\documentstyle[preprint,pra,aps]{revtex}

\begin{document}
\draft
\preprint{RU9583}
\title{Superdeformation in $^{198}$Po}

\author{ D.P.~McNabb,$^{1}$
G.~Baldsiefen,$^{1}$
L.A.~Bernstein,$^{1,2}$
J.A.~Cizewski,$^{1}$
H.-Q.~Jin,\thanks{Present address: Physics Division, Oak Ridge
National Laboratory, Oak Ridge, TN 37831.}$^{1}$
W.~Younes,$^{1}$
J.A.~Becker,$^{2}$
L.P.~Farris,$^{2}$
E.A.~Henry,$^{2}$
J.R.~Hughes,$^{2}$
C.S.~Lee,$^{3}$
S.J.~Asztalos,$^{4}$
B.~Cederwall,\thanks{Present address: Department of Physics, Royal
Institute of Technology,
Stockholm, Sweden.}$^{4}$
R.M.~Clark,$^{4}$
M.A.~Deleplanque,$^{4}$
R.M.~Diamond,$^{4}$
P.~Fallon,$^{4}$
I.Y.~Lee,$^{4}$
A.O.~Macchiavelli,$^{4}$
and F.S.~Stephens$^{4}$}

\address {$^{1}$Department of Physics and Astronomy, Rutgers
University, New Brunswick, NJ 08903}

\address{$^{2}$Lawrence Livermore National Laboratory, Livermore, CA
94550}

\address{$^{3}$Department of Physics, Chung-Ang University, Seoul
156-756, Republic of Korea}

\address {$^{4}$Nuclear Science Division, Lawrence Berkeley National
Laboratory, Berkeley, CA 94720}

\date{\today}
\maketitle

\begin{abstract}
The $^{174}$Yb($^{29}$Si,5n) reaction at 148 MeV with thin targets was
used to populate high-angular momentum states in $^{198}$Po. Resulting
$\gamma$ rays were observed with Gammasphere.  A weakly-populated
superdeformed band of 10 $\gamma$-ray transitions was found and has
been assigned to $^{198}$Po.  This is the first observation of a SD
band in the $A \approx 190$ region in a nucleus with $Z > 83$.  The
${\cal J}^{(2)}$ of the new band is very similar to those of the yrast
SD bands in $^{194}$Hg and $^{196}$Pb.  The intensity profile suggests
that this band is populated through states close to where the SD band
crosses the yrast line and the angular momentum at which the fission
process dominates.

\end{abstract}

\narrowtext

More than 40 superdeformed (SD) bands have been identified to date in
the $A \approx 190$ region.  They have in common characteristics which
include (1) a $\gamma$-ray energy spacing which results in an
upsloping dynamic moment of inertia, ${\cal J}^{(2)}$, with respect to
rotational frequency, and (2) an intensity pattern that suggests SD
bands are populated over several of the highest-spin states and are
sharply depopulated over 1--3 of the lowest energy states.  The
population is thought to occur in the region at which the SD band
becomes yrast~\cite{schiffer89,lauritsen92}.

Mapping the existence of SD bands in the $A \approx 190$ region and
their properties as a function of neutron and proton number is
important for understanding the nuclear structure effects that lead to
the development of the second well.  Three SD bands have recently been
observed in bismuth (Z=83) nuclei~\cite{clark95}.  Theoretical
calculations of Po nuclei (e.g.~\cite{krieger92,satula91}) suggest
that the second well in $^{198}$Po exists at an excitation energy of
about 4 MeV and a well depth of about 2 MeV at $I = 0$.  However,
a simple comparison of the fissility parameter, $Z^2/A$, indicates
that $^{198}$Po is much more prone to fission than other nuclei
studied in this region.  Studies of superdeformation in polonium are
therefore difficult because fission contributes to a larger
spectroscopic background and reduces the population of high angular
momentum states that populate SD bands.  The measurement reported here
shows evidence for a SD band in $^{198}$Po (Z=84), the largest proton
number for which a SD band in the $A \approx 190$ region has been
found.

In our initial experiment~\cite{mcnabb94} with Gammasphere Early
Implementation, a candidate for a SD band was found.  At that time,
the data did not have sufficient statistics to determine whether the
transitions were actually part of a SD cascade.  Here we report the
results from the second experiment which confirm the existence of the
band.

We observed $^{198}$Po with the $^{174}$Yb($^{29}$Si,5n) reaction at a
beam energy of 148 MeV\@.  The beam was provided by the 88-Inch
Cyclotron Facility at the Lawrence Berkeley National Laboratory.  The
target consisted of three, self-supporting enriched $^{174}$Yb
($\approx 98\%$) foils each with a thickness of 600 $\mu$g cm$^{-2}$.
The $\ell_{\rm MAX}$ for this reaction in the middle of the target is
$\sim 38\hbar$~\cite{pace}.  The $\gamma$-ray spectroscopy was done
with the Gammasphere array which, at the time of the experiment,
consisted of 56 Compton-suppressed Ge detectors with Ta-Cu absorbers
placed in front to reduce x-rays.  A total of $1.4 \times 10^9$
three- and higher-fold coincidence events was collected.  The two most
intensely populated evaporation residue channels were $^{198}$Po and
$^{199}$Po ($\approx 50\%$ and $\approx 40\%$, respectively).
Unfolded triples were sorted into a symmetrized
$\gamma$-$\gamma$-$\gamma$ cube with $\gamma$-ray energies ranging
from 100 to 767 keV.  The sum spectrum produced by all combinations of
uncontaminated double gates on the band transitions is shown in
Fig.~1a.  As is evident from the figure, this cascade is rather weak,
and therefore, analysis is prone to background problems.  To minimize
background effects we used higher-fold data in sorting a
$\gamma$-$\gamma$ matrix which was double gated on all combinations of
$\gamma$-ray energies in the cascade.  The spectrum of the SD band
shown in Fig.~1b was generated by gating on the band transitions in
this matrix and subtracting a background spectrum obtained from
selected background gates.

The analysis of these data produced a spectrum with a $\gamma$-ray
sequence that has properties characteristic of SD bands in this mass
region.  Limited statistics prevented a Directional Correlations of
Oriented nuclei (DCO) analysis to establish transition multipolarity.
In this paper we assume that this is a SD structure of E2 transitions.
The search algorithm developed by Hughes, $et~al.$~\cite{hughes94} was
used to aid in the process of looking for additional SD bands.  No
other cascade with energy spacings characteristic of a SD band was
found in the data set.

We have assigned the SD band to $^{198}$Po because the spectrum in
Fig.~1b clearly indicates that this cascade is in coincidence with low
excitation energy transitions in $^{198}$Po, including the 559.2-keV
[$6^+_1 \rightarrow 4^+_1$], 553.2-keV [$4^+_1 \rightarrow 2^+_1$],
and 604.6-keV [$2^+_1 \rightarrow 0^+_1$] transitions~\cite{ensdf}.
It should be noted, however, that the gates on the 307.3- and
429.5-keV band members are possibly contaminated by the 305.8-keV
[$7^- \rightarrow 5^-$] and the 428.4-keV [$6^+_3 \rightarrow 5^-$]
transitions in $^{198}$Po.  However, when these possible contaminants
are removed from the gating conditions, the 604.6 and 553.2-keV lines
persist.  Thus, the low-lying lines are due to the decay of the SD
band to the first well states of $^{198}$Po.  In addition, we see no
evidence of proton emission channels, which would form Bi nuclei.
There are some $\alpha$-channel events, mainly to $^{194-196}$Pb.
However, there is no evidence for the SD bands previously
observed~\cite{brinkman90,theine90,hubel90,farris95,wang91,moore93,clark94}
in
these nuclei, which provides further support of the $^{198}$Po
assignment.

The transition energies and relative intensities, given in Table~I,
were obtained from a sum of all combinations of double gates of the SD
$\gamma$-ray energies in the sorted cube.  A generalized background
subtraction~\cite{radford95} was employed for these spectra.  We
estimate the upper limit of the relative intensity of the SD band
compared with the normal states to be 0.3\%.  This was obtained by
dividing the intensity of the 307.3-keV SD transition, as seen in the
spectrum double gated on 349.3- and 390.5-keV transitions, by the
summed intensity of transitions feeding the $0^+_1$ ground state level
and the known 750-ns 12$^+$ isomer in $^{198}$Po.

The $^{198}$Po SD band has properties characteristic of most SD bands
found in the $A \approx 190$ region.  The average spacing of the
$\gamma$-ray energies is $\overline{\Delta E_{\gamma}} = 40.7(5)$ keV
for frequencies between $\hbar\omega$ = 0.088 and 0.271 MeV.  The
dynamic moment of inertia for this band is displayed in Fig.~2, where
it is compared with selected SD bands in $A \approx 190$ even-even
nuclei.  It is interesting to note that the ${\cal J}^{(2)}$ of the
$^{198}$Po band is most similar to the SD yrast bands in
$^{194}$Hg~\cite{riley90,cederwall94} and
$^{196}$Pb~\cite{wang91,moore93,clark94}.  The rise in the ${\cal
J}^{(2)}$ for
these three isotones is understood as the gradual alignment of pairs
of $j_{15/2}$ neutrons and $i_{13/2}$ protons under the influence of
weak pairing correlations~\cite{riley90}.  The small differences of
magnitude for the ${\cal J}^{(2)}$ can possibly be attributed to
differences in pairing and/or deformation, but the calculations are
not sufficiently sensitive to predict such small effects.

Using the method described by Becker $et~al.$~\cite{becker92}, the
fitted spin of the level populated by the 175.9-keV transition is
6.1(1)$\hbar$.  This near integer value for the spin lends further
credence to assigning the band to the even-even $^{198}$Po.  This spin
assignment is also consistent with the average spin $\approx 4\hbar$
of the low-excitation energy states populated by the SD band decay.

An interesting feature of this SD band is that the highest observed SD
transition has an energy below 550 keV, which corresponds to a
relatively low spin transition, $26\hbar\rightarrow 24\hbar$.  Most
yrast SD bands in A$\approx$190 nuclei extend to transition energies
above 650 keV\@.  Our analysis of the population of the normal states
in $^{198}$Po~\cite{mcnabb94} determined that fission dominates
compound nucleus decay above about 22$\hbar$.  A reasonable
explanation for not observing any $\gamma$-ray transition in the SD
band above 550 keV in $^{198}$Po is that the fission process strongly
dominates above $\approx 26\hbar$ --- population of even
highly-deformed configurations in the evaporation residue cannot
compete with fission at higher angular momentum.

The intensity profile, shown in Fig.~1a, indicates that the SD band is
sharply fed over few transitions.  The fission process is limiting the
maximum angular momentum of the $^{198}$Po residue, thereby reducing
the entry region over which the SD band is populated.  Since the
population of SD bands is thought to occur in the region where the SD
band crosses the yrast line~\cite{schiffer89,lauritsen92}, the
intensity analysis suggests that the SD band is crossing the yrast
line at roughly the same spin at which fission begins to dominate the
$^{198}$Po channel.  The rather low intensity of the band relative to
the total intensity in the $5n$ channel supports the conclusion that
we are close to the limit where the fission process cuts off the entry
region, which is dependent upon the spin at which the SD band becomes
yrast.

To summarize, we have observed a new SD band which has been assigned
to $^{198}$Po.  Our results represent the first observation of a
superdeformed band in the $A \approx 190$ region with $Z>83$.  The
upsloping ${\cal J}^{(2)}$ is very similar to that observed in other
N=114 isotones, most likely reflecting the role of aligning $j_{15/2}$
neutrons.  The population of this band and its intensity profile
suggests that $^{198}$Po is close to the limit where the fission
process cuts off the entry region to the SD band.  Observation of a SD
configuration in $^{198}$Po, in spite of intense competition from
fission, suggests that this region of superdeformed shapes can be
extended to heavier nuclei where SD minima are predicted to exist.
However, a successful search for SD excitations in $Z>84$ systems will
require high statistics data, such as will be available with full
implementation Gammasphere.

This work has been funded in part by the National Science Foundation
(Rutgers), the U.S. Department of Energy, under contracts No.
W-7405-ENG-48 (LLNL), and AC03-76SF00098 (LBL), and the Basic Science
Research Institute Program, Ministry of Education, Korea (Project
No. BSRI-95-2417).

\begin{figure}
\caption{Spectra of the $^{198}$Po SD band with different sorting
conditions.  (a) Generated by summing all combinations of clean double
gates on the SD transitions listed in Table~I.  The inset shows the
intensity distribution of these SD transitions.  (b) Generated by
summing all combinations of triple gates on the SD transitions listed
in Table~I.}
\label{fig-spec}
\end{figure}

\begin{figure}
\caption[]{Comparison of ${\cal J}^{(2)}$ of the $^{198}$Po SD band with
SD bands in $^{196}$Pb and $^{194}$Hg.  Data are taken from
Refs.~\cite{clark94,cederwall94} and the present work.}
\label{fig-j2}
\end{figure}

\begin{table}
\caption{Energies and relative intensities of $^{198}$Po SD transitions.}
\label{tab-gamenergies}
\begin{tabular}{ccdc}
& Energy(keV)&Intensity\tablenotemark[1] & \\
\tableline
 & 175.91(27) & -\tablenotemark[2] & \\
 & 220.37(20) & 0.47(8) & \\
 & 264.59(15) & 0.85(9) & \\
 & 307.29(15) & 0.97(9) & \\
 & 349.29(16) & 1.00(10) & \\
 & 390.46(20) & 0.89(10) & \\
 & 429.54(19) & -\tablenotemark[2] & \\
 & 467.90(38) & 0.87(10) & \\
 & 505.85(42) & 0.87(12) & \\
 & 542.57(42) & 0.40(9) & \\
\end{tabular}

\tablenotetext[1]{Intensities have been corrected for
detector efficiency and electron conversion.}
\tablenotetext[2]{The intensities of these transitions have not been
extracted.  The contaminants are a 176-keV line from the Coulomb
excitation of the target and a low-lying 429-keV transition in
$^{198}$Po~\cite{mcnabb94}.}
\end{table}
\end{document}